\documentclass[twocolumn,article,amsmath,amssymb,aps]{revtex4}
\usepackage{graphicx}
\usepackage{epsfig}
\usepackage{multirow}
\begin{document}

\title{New features of the triaxial nuclei described with a coherent state model }

\author{A. A. Raduta$^{a),b)}$, P. Buganu$^{a)}$ and Amand Faessler $^{c)}$}

\address{$^{a)}$ Department of Theoretical Physics, Institute of Physics and
  Nuclear Engineering,POBox MG6, Bucharest 077125, Romania}

\address{$^{b)}$Academy of Romanian Scientists, 54 Splaiul Independentei, Bucharest 050094, Romania}

\address{$^{c)}$Instit\"{u}t f\"{u}r Theoretishe Physik der Universit\"{a}t T\"{u}bingen, Auf der Morgenstelle 14, D-72076 T\"{u}bingen,  Germany}

\begin{abstract} Supplementing  the Liquid Drop Model (LDM) Hamiltonian, written in the intrinsic reference frame, with a sextic oscillator plus a centrifugal term in the variable $\beta$ and  a potential in $\gamma$ with a minimum in $\frac{\pi}{6}$, the Sch\"{o}dinger equation is separated  for the two variables which results in having a new description for the triaxial nuclei, called Sextic and Mathieu Approach (SMA). SMA is applied for two non-axial nuclei, $^{180}$Hf and $^{182}$W and results are compared with those yielded by the Coherent State Model (CSM). As the main result of this paper we derive analytically the equations characterizing SMA from a semi-classical treatment of the CSM Hamiltonian. In this manner the potentials in $\beta$ and $\gamma$ variables respectively, show up in a quite natural way which contrasts their  ad-hoc choice when SMA emerges from LDM.    
\end{abstract}

\maketitle

\renewcommand{\theequation}{1.\arabic{equation}}
\setcounter{equation}{0}
\section{Introduction}
\label{sec:level1}
Many properties of the low lying spectra of even-even nuclei can be described in terms of specific dynamic symmetries \cite{Iache,Iache1} associated with a definite nuclear phase. The transition from one symmetry to another is therefore interpreted as a phase transition. On the path of such a transition a critical point is met.
The spectroscopic properties of the nuclei corresponding to the critical points are in general difficult to be described.

In Ref. \cite{Gino,Diep}, it has been proved
that on the transition from the $U(5)$ to the $O(6)$ symmetry a critical point exists
for a second order phase transition while the transition from
the $U(5)$ to the $SU(3)$ symmetry has the features of  a first order phase transition. In Ref.\cite{McCutchan1} it was proved that most of nuclei  are mapped  not on the border of the symmetry triangle introduced by Casten \cite{Cast} but in the interior of the triangle. Examples of such nuclei are the $Os$ isotopes \cite{McCutchan2}.   

Recently, Iachello \cite{Iache2,Iache9} pointed out that the mentioned  critical points
correspond to
distinct symmetries, namely $E(5)$ and $X(5)$, respectively. For the critical value
of an
ordering parameter, energies are given by the zeros of a Bessel function of half integer and irrational indices, respectively \cite{Zam,Zam1,Biz1}. In Ref.\cite{Biz2} the X(5) description was extended to the first octupole vibrational band in nuclei close to axial symmetry and also close to the critical point of the U(5) to SU(3) phase transition. 
Another symmetries, called $Y(5)$ and $Z(5)$,  have been pointed out
in Refs.\cite{Iache3,Bona1}. The former symmetry corresponds to the critical point of the transition from axial to triaxial nuclei while the latter one is related to the critical point of the transition from prolate to oblate  through a triaxial shape.

The nice feature of the critical point symmetry is that the description in the intrinsic frame
is performed by two separated differential equations for beta and gamma degrees of freedom.
These equations are solvable and the solutions are irreducible representations for the specific symmetry. Moreover, apart from an overall  scaling parameter the energies are parameter free quantities. Since the idea of symmetries associated to the critical points of various phase transitions showed up, many attempts have been made to describe the two dynamic deformations by
solvable and separable differential equations with specific beta and gamma potentials. Since the triaxial nuclei might be considered as critical points for a phase transition from prolate to oblate shape one expects that they can be described by specific solvable models. Thus, a description of soft gamma nuclei around $\gamma^0=\frac{\pi}{6}$ with an oscillator potential in $\gamma$ and a Kratzer potential in $\beta$ has been developed in Refs.\cite{fortu1,fortu2,fortu3}.

Another model was proposed by two of us (A.A.R. and P.B.) in a previous publication \cite{RaBu011}. Indeed, the Liquid Drop Model
(LDM) Hamiltonian \cite{Bo} written in the intrinsic frame is separated into two terms describing the $\beta$ and $\gamma$ variables.
The potential in $\beta$ consists in a centrifugal term and a sextic oscillator potential,  while the differential equation for $\gamma$ is that for the Mathieu function. Due to this feature we  called the formalism developed there as the {\bf S}extic and {\bf M}athieu {\bf A}pproach (SMA).
The proposed model was applied for five nuclei which exhibit the signature for triaxiality, $^{188}$Os, $^{190}$Os, $^{192}$Os, $^{228}$Th,
$^{230}$Th.    

Here we continue the study of SMA by describing another two nonaxial isotopes, $^{180}$Hf and $^{182}$W. In Ref.\cite{RaBu011} we remarked that the agreements with experimental data provided by SMA and the Coherent State Model (CSM) respectively, are of similar quality. Here we attempt  to answer the question {\it is there any reason for that to happen?}

The objectives of this paper are described according to the following plan. In Section II, a brief review of the main ingredients defining SMA is presented.
Section III is devoted to the CSM approach. It is shown, in Section IV, that SMA's equations are obtainable from CSM under certain circumstances.
The numerical application is discussed in Section V, while the conclusions are drawn in Section VI.

\renewcommand{\theequation}{2.\arabic{equation}}
\setcounter{equation}{0}
\section{Sextic and Mathieu Approach }
\label{sec:level2}
 Adding to the Bohr-Mottelson Hamiltonian written in the intrinsic frame of reference \cite{Bo} a potential of a convenient form:
\begin{equation}
V(\beta,\gamma)=V_1(\beta)+\frac{1}{\beta^2}V_2(\gamma)
\end{equation}
the equations for the variables $\beta$ and $\gamma$ are separated from each other. Thus, the equation for $\beta$ reads:
\begin{equation}
\left[-\frac{1}{\beta^{4}}\frac{\partial}{\partial\beta}\beta^{4}\frac{\partial}{\partial\beta}+\frac{L(L+1)}{\beta^{2}}+v_{1}(\beta)\right]f(\beta)=\varepsilon_{\beta}f(\beta),
\label{eqbeta}
\end{equation}
The remaining terms, depend on $\gamma$ but also on $\beta$ by means of the factor
 $1/\beta^{2}$. In order that the variable separation is achieved,  the mentioned factor is replaced by an average value $1/\langle\beta^{2}\rangle$. Actually, in our concrete calculation this is considered to be a free parameter.
The resulting equation in  gamma variable, is:
\begin{eqnarray}
&&\left[-\frac{1}{\sin3\gamma}\frac{\partial}{\partial\gamma}\sin3\gamma\frac{\partial}{\partial\gamma}-\frac{3}{4}R^{2}+\left(10L(L+1)-\frac{39}{4}R^{2}\right)\right.\nonumber\\
&&\times\left.\left(\gamma-\frac{\pi}{6}\right)^{2}+v_{2}(\gamma)\right]\phi(\gamma)=\tilde{\varepsilon}_{\gamma}\phi(\gamma),
\label{eqgamma}
\end{eqnarray}
where the following notations are used:
\begin{eqnarray}
v_{1}(\beta)&=&\frac{2B}{\hbar^{2}}V_{1}(\beta),\hspace{0.5cm}v_{2}(\gamma)=\frac{2B}{\hbar^{2}}V_{2}(\gamma),\nonumber\\
\varepsilon_{\beta}&=&\frac{2B}{\hbar^{2}}E_{\beta},\hspace{0.5cm}\tilde{\varepsilon}_{\gamma}=\langle\beta^{2}\rangle\frac{2B}{\hbar^{2}}E_{\gamma}.
\label{enreduse}
\end{eqnarray}
In order to solve the separated equations in $\beta$ and $\gamma$ respectively, we have to specify the potentials $v_1(\beta)$ and $v_2(\gamma)$.
$v_1(\beta)$ is a sextic oscillator potential in $\beta$ and the corresponding differential equation is quasi-exactly solvable.
Indeed, by changing the function $f(\beta)=\beta^{-2}\varphi(\beta)$ in Eq.(\ref{eqbeta}) one obtains:
\begin{equation}
\left[-\frac{\partial^{2}}{\partial\beta^{2}}+\frac{L(L+1)+2}{\beta^{2}}+v_{1}(\beta)\right]\varphi(\beta)=\varepsilon_{\beta}\varphi(\beta).
\label{newbetaeq}
\end{equation}

One can check that this equation can be identified to the Schr\"{o}dinger equation
\begin{equation}
H_{x}\psi(x)=E\psi(x)
\label{Hasx}
\end{equation}
associated to a sextic oscillator with a centrifugal barrier
\begin{eqnarray}
&&H_{x}=-\frac{\partial^{2}}{\partial x^{2}}+\frac{\left(2s-\frac{1}{2}\right)\left(2s-\frac{3}{2}\right)}{x^{2}}\\
&&+[b^{2}-4a(s+\frac{1}{2}+M)]x^{2}
+2abx^{4}+a^{2}x^{6},\nonumber
\label{sexticpot}
\end{eqnarray}
if the following correspondence is adopted:
\begin{eqnarray}
&&x=\beta,\hspace{0.2cm}E=\varepsilon_{\beta},\hspace{0.2cm}\left(2s-\frac{1}{2}\right)\left(2s-\frac{3}{2}\right)=L(L+1),\nonumber\\
&&\hspace{0.4cm}v_{1}(\beta)=(b^{2}-4ac)\beta^{2}+2ab\beta^{4}+a^{2}\beta^{6},\nonumber\\
&&s=\frac{L}{2}+\frac{3}{4},\hspace{0.2cm}c=\frac{L}{2}+\frac{5}{4}+M.
\label{v1debeta}
\end{eqnarray}
The eigenfunctions provided by Eq.(\ref{newbetaeq}) are written in the form:
\begin{eqnarray}
&&\varphi_{n_{\beta},L}^{(M)}(\beta)=N_{n_{\beta},L}P_{n_{\beta},L}^{(M)}(\beta^{2})\beta^{2s-\frac{1}{2}}e^{-\frac{a}{4}\beta^{4}-\frac{b}{2}\beta^{2}},
\nonumber\\
&&n_{\beta}=0,1,2,...M,
\end{eqnarray}
where $N_{n_{\beta},L}$ are normalization constants and $P_{n_{\beta},L}^{(M)}(\beta^{2})$
are polynomials of degree $n_{\beta}$ in $\beta^2$, whose coefficients form an (M+1)-vector which satisfies an eigenvalue equation, the corresponding eigenvalue being denoted by 
$\lambda^{(M)}_{n_{\beta}}(L)$. Using the notations from Eq.(\ref{enreduse}), one obtains for the eigenvalues the following expression: 
\begin{eqnarray}
&&E_{\beta}(n_{\beta},L)=\frac{\hbar^{2}}{2B}\left[4bs(L)+\lambda_{n_{\beta}}^{(M)}(L)+u_{0}^{\pi}\right],\nonumber\\
&&n_{\beta}=0,1,2,...,M.
\label{energybe}
\end{eqnarray}
Here $u_{0}^{\pi}$ denotes two constants to  be fixed such that for the minima $(\beta_{min}^{\pi}>0)$ of the  potentials $v_{1}^{+}(\beta)$ and $v_{1}^{-}(\beta)$ given by
\begin{equation}
v_{1}^{\pi}(\beta)=(b^{2}-4ac^{\pi})\beta^{2}+2ab\beta^{4}+a^{2}\beta^{6}+u_{0}^{\pi}\hspace{0.2cm}(\pi\equiv\pm),
\end{equation}
to have the same energy. Details about how to solve the eigenvalue equation for a sextic oscillator plus a centrifugal term can be found in Ref.\cite{Ushve}.

Concerning the equation for the variable $\gamma$, Eq.(\ref{eqgamma}) can be reduced to the Mathieu equation \cite{Raduta}. First we change the function
\begin{equation}
\phi(\gamma)=\frac{{\cal M}(3\gamma)}{\sqrt{|\sin3\gamma|}}.
\end{equation}
The equation for the new function is:
\begin{eqnarray}
&&\left[\frac{\partial^{2}}{\partial\gamma^{2}}+\left(\tilde{\varepsilon}_{\gamma}+\frac{1}{4}+\frac{3}{4}R^{2}\right)+\frac{9}{4\sin^{2}3\gamma}
\right.\\
&-&\left.\left(10L(L+1)-\frac{39}{4}R^{2}\right)\left(\gamma-\frac{\pi}{6}\right)^{2}-v_{2}(\gamma)\right]{\cal M}(3\gamma)=0.\nonumber
\end{eqnarray}
where L denotes the angular momentum and R is its projection on the axes OX.
The potential in $\gamma$ is chosen to exhibit a minimum at  $\gamma_{0}=\pi/6$:
\begin{equation}
v_{2}(\gamma)=\mu\cos^{2}3\gamma.
\end{equation}
Making the Taylor expansions around the minimum value of the gamma potential:

\begin{equation}
\frac{9}{4\sin^{2}3\gamma}\sim\frac{9}{4}+\frac{81}{4}\left(\gamma-\frac{\pi}{6}\right)^{2},\hspace{0.2cm}\mu\cos^{2}3\gamma\sim9\mu\left(\gamma-\frac{\pi}{6}\right)^{2},
\end{equation}
the equation for the variable $\gamma$ becomes:
\begin{eqnarray}
&&\left[\frac{\partial^{2}}{\partial\gamma^{2}}-\left(10L(L+1)-\frac{39}{4}R^{2}+9\mu-\frac{81}{4}\right)
\left(\gamma-\frac{\pi}{6}\right)^{2}\right.\nonumber\\
&&\left.+\left(\tilde{\varepsilon}_{\gamma}+\frac{3}{4}R^{2}+\frac{5}{2}\right)\right]{\cal M}(3\gamma)=0.
\label{Mat1}
\end{eqnarray}
Using again in (\ref{Mat1})  the approximation
\begin{equation}
\left(\gamma-\frac{\pi}{6}\right)^{2}\approx \frac{1}{18}(\cos6\gamma +1),
\end{equation}
and making the change of variable $y=3\gamma$, we obtain
\begin{equation}
\left(\frac{\partial^{2}}{\partial y^{2}}+a-2q\cos2y\right){\cal M}(y)=0,
\label{Mat2}
\end{equation}
where
\begin{eqnarray}
&&q=\frac{1}{36}\left(\frac{10}{9}L(L+1)-\frac{13}{12}R^{2}+\mu-\frac{9}{4}\right),\nonumber\\
&&a=\frac{1}{9}\left(\tilde{\varepsilon}_{\gamma}+\frac{3}{4}R^{2}+\frac{5}{2}\right)-2q.
\label{qanda}
\end{eqnarray}
Eq. (\ref{Mat2}) is just the well known Mathieu equation.
Using the expression for the characteristic value $a$, Eq.(\ref{qanda}), of the Mathieu equation one can find the expression for the excitation energy of the $\gamma$ equation
\begin{eqnarray}
&&E_{\gamma}(n_{\gamma},L,R)=\frac{\hbar^{2}}{2B}\frac{1}{\langle\beta^{2}\rangle}\nonumber\\
&\times&\left[9a_{n_{\gamma}}(L,R)+18q(L,R)-\frac{3}{4}R^{2}-\frac{5}{2}\right],\nonumber\\
&&n_{\gamma}=0,1,2,....
\label{energyga}
\end{eqnarray}

The total energy for the  system  is obtained by adding the energies given by the equations  (\ref{energybe}) and (\ref{energyga}):
\begin{equation}
E(n_{\beta},n_{\gamma},L,R)=E_{0}+E_{\beta}(n_{\beta},L)+E_{\gamma}(n_{\gamma},L,R)
\label{totalen}
\end{equation}
The excitation energies depend on four quantum numbers, $n_{\beta}$, $n_{\gamma}$, $L$, $R$, and five parameters $\hbar^{2}/2B$, $a$, $b$, $\frac{1}{\langle\beta^{2}\rangle}$, $\mu$.

The quantum numbers defining the ground, beta and gamma bands are as follows:
\begin{eqnarray}
n_{\beta}&=&0,~n_{\gamma}=0,\hspace{0.5cm}R=L,\hspace{0.7cm}L=0,2,4,...~\rm{g~~band},\nonumber\\
n_{\beta}&=&0,~n_{\gamma}=1,~\Bigg\{{{R=L-2, L=2,4,6,...} \atop {R=L-1, L=3,5,7,...}}~\gamma ~ \rm{band},\nonumber\\
n_{\beta}&=&1,~n_{\gamma}=0,\;R=L,\;\;L=0,2,4,...~\beta~\rm{band}.\nonumber\\
\end{eqnarray}.

The wave function describing the whole system is:
\begin{eqnarray}
&&|LRMn_{\beta}n_{\gamma}\rangle=N_{L,n_{\beta}}N_{L,R,n_{\gamma}}f_{L,n_{\beta}}(\beta)\phi_{L,R,n_{\gamma}}(\gamma)\nonumber\\
&\times&\sqrt{\frac{2L+1}{16\pi^{2}(1+\delta_{R0})}}\left(D_{MR}^{L}(\Omega)+(-1)^{L}D_{M-R}^{L}(\Omega)\right).\nonumber\\
\end{eqnarray}
where the factors $N_{L,n_{\beta}}$ and $N_{L,R,n_{\gamma}}$ denote the norms of the partial wave functions.

Note that when the matrix elements with the wave functions depending on $\beta$ and $\gamma$ respectively are calculated, the integration over the $\beta$ is performed with the measure $\beta^{4}d\beta$, while that over the $\gamma$  with the measure $|\sin3\gamma|d\gamma$. These measures are characterizing the $(\beta,\gamma)$ space within the liquid drop model.
These wave functions are further used to calculate the reduced E2 transition probabilities.

In our approach the quadrupole transition operator is defined as:
\begin{eqnarray}
&&T_{2\mu}^{(E2)}=t_1\beta \left[\cos\left(\gamma-\frac{2\pi}{3}\right) D^2_{\mu 0}\right.\nonumber\\
&+&\left.\frac{1}{\sqrt{2}}\sin\left(\gamma-\frac{2\pi}{3}\right)(D^2_{\mu 2}+D^2_{\mu, -2})\right]\nonumber\\
&+ &t_2\sqrt{\frac{2}{7}}\beta^2 \left[-\cos\left(2\gamma-\frac{4\pi}{3}\right) D^2_{\mu 0}\right.\nonumber\\
&+&\left.\frac{1}{\sqrt{2}}\sin\left(2\gamma-\frac{4\pi}{3}\right)(D^2_{\mu 2}+D^2_{\mu, -2})\right].
\label{tranoper}
\end{eqnarray}
The argument $\gamma-2\pi/3$ of the trigonometric functions is justified by the fact that it defines the axes 1 of the principal inertial ellipsoid. Indeed, the transformation from the laboratory to the intrinsic frame is a rotation defined by the matrix $D^L_{MR}$ where the quantum numbers $M$ and $R$ are eigenvalues of the operator $Q_1$.

The reduced E2 transition probabilities are defined as:

\begin{equation}
B(E2,J_{i}\rightarrow J_{f})=|\langle J_{i}||T_{2}^{(E2)}||J_{f}\rangle|^{2}.
\end{equation}
where the Rose's convention \cite{Rose} was used for the reduced matrix elements.

Summarizing, the SMA formalism uses a sextic oscillator potential with a centrifugal term for the $\beta$ and a Mathieu equation for the $\gamma$ variable. These equations provide for the total energy given by Eq.(\ref{totalen}) a compact form. The wave functions obtained by solving the quoted equations together with the transition operator of Eq.(\ref{tranoper}), are used to calculate the electric quadrupole transition probabilities. 

There are several groups which studied the $\gamma$ soft nuclei around $\gamma^0=\frac{\pi}{6}$
\cite{Bona1,fortu1,fortu2,fortu3}. The quoted approaches differ from the present formalism by the 
equations used for the description of $\beta$ and $\gamma$ coordinates.

Since the results of the SMA formalism will be compared with those obtained by CSM, in what follows we shall briefly present the main ingredients of the latter approach.

\renewcommand{\theequation}{3.\arabic{equation}}
\setcounter{equation}{0}
\section{Coherent state model (CSM)}
\label{sec:level 3}
CSM defines \cite{Rad1} first a restricted collective space whose vectors  are
model states of ground, $\beta$ and $\gamma$ bands. In choosing these states we
were guided by some experimental information which results in formulating a set of criteria to be fulfilled by the searched states.

All these restrictions required are fulfilled by the following set of three deformed quadrupole boson states:

\begin{equation}
\psi_g=e^{[d(b^{\dagger}_0-b_0)]}|0\rangle\equiv T|0\rangle,~
\psi_{\gamma}=\Omega^{\dagger}_{\gamma,2}\psi_g,~
\psi_{\beta}=\Omega^{\dagger}_{\beta}\psi_g.
\label{psigbga}
\end{equation}

where the excitation operators for $\beta$ and $\gamma$ bands are defined by:
\begin{eqnarray}
&&\Omega^{\dagger}_{\gamma,2}=(b^{\dagger}b^{\dagger})_{2,2}+d\sqrt{\frac{2}{7}}
b^{\dagger}_{2,2}, \nonumber\\
&&\Omega^{\dagger}_{\beta}=(b^{\dagger}b^{\dagger}b^{\dagger})_0 +\frac{3d}{\sqrt{14}}
(b^{\dagger}b^{\dagger})_0 -\frac{d^3}{\sqrt{70}}.
\label{omegabe}
\end{eqnarray}

From the three deformed states one generates through projection, three sets of
mutually orthogonal states
\begin{equation}
\varphi^i_{JM}=N^i_JP^J_{M0}\psi_i, i=g,\beta,\gamma,
\label{projfi}
\end{equation}
where $P^J_{MK}$ denotes the projection operator:

\begin{equation}
P^J_{MK}=\frac{2J+1}{8\pi^2}\int {D^{J^*}_{MK}\hat{R}(\Omega)d\Omega},
\label{projop}
\end{equation}
and $N^i_{J}$  the normalization factors and $D^J_{MK}$ the rotation matrix elements.
It was proved that the deformed and projected states contain the salient features of the major collective bands.
Since we attempt to set up a very simple model we relay on the experimental feature saying that the $\beta$  band is largely decoupled from the ground as well as from the $\gamma$ bands and choose a model Hamiltonian  whose matrix elements
between  beta states and  states belonging either to the ground or to the gamma band are all equal to zero.
The simplest Hamiltonian obeying this restriction is
\begin{equation}
H=A_1(22\hat{N}+5\Omega^{\dag}_{\beta'}\Omega_{\beta'})+A_2\hat{J}^2
+A_3\Omega^{\dagger}_{\beta}\Omega_{\beta},
\label{hascsm}
\end{equation}
where $\hat{N}$ is the boson number, $\hat{J}^2$-angular momentum squared and $
\Omega^{\dagger}_{\beta'}$ denotes:
\begin{equation}
\Omega^{\dagger}_{\beta'}=(b^{\dagger}b^{\dagger})_{00}-\frac{d^2}{\sqrt{5}}.
\end{equation}

Higher order terms in boson operators can be added to the Hamiltonian $H$ without altering the decoupling condition for the beta band. An example of this kind is the correction:
\begin{equation}
\Delta H=A_4(\Omega^{\dagger}_{\beta}\Omega^2_{\beta^{\prime}}+h.c.)+
A_5\Omega^{\dagger 2}_{\beta^{\prime}}\Omega^2_{\beta^{\prime}}.
\label{deltahascsm}
\end{equation}

The energies for beta band as well as for the gamma band states of odd
angular momentum are described as average values of H (\ref{hascsm}), or $H+\Delta H$ on $\varphi^{\beta}_{JM}$
and $\varphi^{\gamma}_{JM}$ (J-odd), respectively. As for the energies for the ground band and
those of gamma band states  with even angular momentum, they are obtained by diagonalizing
a 2x2 matrix for each J.

The quadrupole transition operator is considered to be a sum of a linear  term in bosons and one which is quadratic in the quadrupole bosons:
\begin{equation}
Q_{2\mu}=q_1(b^{\dag}_{2\mu}+(-)^{\mu}b_{2,-\mu})+
q_2((b^{\dag}b^{\dag})_{2\mu}+(bb)_{2\mu}).
\label{q2anh}
\end{equation}
The form of the anharmonic component of $Q_{2\mu}$ is justified by the fact that this is the lowest
order boson term which may connect the states from beta and ground bands in the vibrational limit, i.e. $d$-small.

Using the Rose convention \cite{Rose}, the reduced probability for the E2 transition $J^+_i\to J^+_f$ can be expressed as:
\begin{equation}
B(E2;J^+_i\to J^+_f)=\left(\langle J^+_i||Q_2||J^+_f\rangle\right)^2
\end{equation}
Three specific features of CSM are worth to be mentioned:

a) The model states are generated through projection from a coherent
state and two excitations of that through simple  polynomial boson operators.
Thus,
it is expected that the projected states may account for the semiclassical
behavior of the nuclear system staying in a state of high spin.

b) The states are infinite series of bosons and thus highly deformed
states can be described.

c) The model Hamiltonian is not commuting with the boson number operator and 
because of this property a basis generated from a coherent state is expected
to be most suitable. 

The CSM has been successfully applied to several nuclei
exhibiting various equilibrium shapes which according to the IBA (Interacting Boson Approximation) classification,
exhibit the  SO(6), SU(5) and SU(3) symmetries, respectively.
Several improvements of CSM has been proposed by considering additional
degrees of freedom
like isospin \cite{Rad2}, quasiparticle \cite{Rad3} or collective octupole coordinates 
\cite{Rad4,RaSa}. CSM has been also used to describe some nonaxial nuclei \cite{RadUve} and the results were compared with those obtained with the Rotation-Vibration Model \cite{GrFa}.
A review of the CSM achievements is found in Ref. \cite{Rad5}.

\renewcommand{\theequation}{4.\arabic{equation}}
\setcounter{equation}{0}
\section{SMA formalism obtained by quantizing the classical CSM equations.}
\label{sec:level4}

In our previous publication on this subject \cite{RaBu011} we were noting that the two approaches SMA and CSM
describe the data of the considered nuclei, equally well. This amazing feature raised the question {\it why that happens?}  Actually, here we aim at answering the mentioned question.
{\it In brief we shall prove that, indeed, SMA can be analytically derived from the CSM formalism.}
The source of our inspiration was the result from Ref.\cite{RaBu} showing that a generalized Davidson potential \cite{Davidson} can be obtained by a semiclassical treatment of a fourth order boson Hamiltonian.

For this purpose we shall study the classical properties emerging from CSM, by dequantizing the specific boson Hamiltonian used by CSM, considering its average with the coherent state:
\begin{equation}
|\psi\rangle=\exp\left[z_{0}b_{0}^{\dagger}+z_{2}b_{2}^{\dagger}+z_{-2}b_{-2}^{\dagger}-z_{0}^{*}b_{0}-z_{2}^{*}b_{2}-z_{-2}^{*}b_{-2}\right]|0\rangle,
\end{equation}
where $z_{k},z_{k}^{*}$ with $k=0,\pm 2$ are complex numbers depending on time. As usual, $|0\rangle$ denotes the vacuum state for the quadrupole bosons.
The basic property of this function is comprised by the equation:
\begin{equation}
b_{\mu}|\psi\rangle=\left(\delta_{\mu,0}z_{0}+\delta_{\mu,2}z_{2}+\delta_{\mu,-2}z_{-2}\right)|\psi\rangle.
\end{equation}
The classical Hamilton function associated to the CSM's model Hamiltonian is:

\begin{eqnarray}
\mathcal{H}&\equiv&\langle\psi|H|\psi\rangle =2(11A_{1}+3A_{2})\left(|z_{0}|^{2}+|z_{2}|^{2}+|z_{-2}|^{2}\right)\nonumber\\
&+&A_{1}\left(2z_{2}^{*}z_{-2}^{*}+z_{0}^{*2}-d^{2}\right)\left(2z_{2}z_{-2}+z_{0}^{2}-d^{2}\right)+\nonumber\\
&+&\frac{A_{3}}{70}\left[2\left(6z_{0}^{*}z_{2}^{*}z_{-2}^{*}-z_{0}^{*3}\right)+3d\left(2z_{2}^{*}z_{-2}^{*}+z_{0}^{*2}\right)-d^{3}\right]\nonumber\\
&\times&\left[2\left(6z_{0}z_{2}z_{-2}-z_{0}^{3}\right)+3d\left(2z_{2}z_{-2}+z_{0}^{2}\right)-d^{3}\right].
\end{eqnarray}
The  equations of motion described by the classical coordinates $z_0, z_{\pm 2}$ and their complex conjugates  $z_0^*, z_{\pm 2}^*$ are obtained from the variational principle of the minimum action:
\begin{equation}
\delta\int_{0}^{t}\langle\psi|H-i\hbar\frac{\partial}{\partial t'}|\psi\rangle dt'=0.
\end{equation}
The action variation can be written in a compact form:
\begin{eqnarray}
\delta\langle \psi|i\hbar\frac{\partial}{\partial t}|\psi\rangle &=&
\stackrel{\cdot}{z}_0\delta z_0^*-\stackrel{\cdot}{z}_0^* \delta z_0\nonumber\\ 
&+&\stackrel{\cdot}{z}_2\delta z_2^*-\stackrel{\cdot}{z}_2^* \delta z_2\nonumber\\
&+&\stackrel{\cdot}{z}_{-2}\delta z_{-2}^*-\stackrel{\cdot}{z}_{-2}^* \delta z_{-2}.
\end{eqnarray}
where the symbol $"\stackrel{\cdot}{}"$ stands for the time derivative.
Finally, the result for the classical equations is:
\begin{eqnarray}
\frac{\partial H}{\partial z_k}&=&-i\stackrel{\cdot}{z}_k^{*},\nonumber\\
\frac{\partial H}{\partial z_k^{*}}&=&i\stackrel{\cdot}{z}_k,\;\;k=0,\pm 2 .
\end{eqnarray}
These equations support the interpretation of $z_k$, with $k=0,\pm 2$, as classical phase space coordinates and of  $z_k^*$ as the corresponding conjugate momenta.
With the complex coordinates we may define the canonical conjugate coordinates:
\begin{eqnarray}
Q_0&=&\frac{z_0+z_0^{*}}{\sqrt{2}},\;Q_2=\frac{z_{-2}+z_2^{*}}{\sqrt{2}},\;
Q_{-2}=\frac{z_2+z_{-2}^{*}}{\sqrt{2}},\nonumber\\
P_0&=&\frac{z_0-z_0^{*}}{\sqrt{2}},\;P_2=\frac{z_2-z_{-2}^{*}}{\sqrt{2}},\;
P_{-2}=\frac{z_2-z_{2}^{*}}{\sqrt{2}},
\end{eqnarray}
which, evidently, obey the equations:
\begin{eqnarray}
\{Q_0,P_0\}&=&1,\;\;\{Q_{\pm},P_{\pm}\}=1,\nonumber\\
\{Q_k,{\cal H}\}&=&\stackrel{\cdot}{Q}_k,\;\;\{P_k,{\cal H}\}=\stackrel{\cdot}{P}_k.
\end{eqnarray}
For two given functions $f,g$ defined on the phase space, their Poisson bracket is denoted by
 $\{f,g\}$ and defined as:
\begin{equation}
\{f,g\} =\sum \left(\frac{\partial f}{\partial Q_k}\frac{\partial g}{\partial P_k}-
\frac{\partial f}{\partial P_k}\frac{\partial g}{\partial Q_k}\right).
\end{equation}
In terms of the canonical coordinates $Q$ and $P$ the classical energy function is given in Appendix A.

In what follows we shall study the Hamilton function ${\cal H}$ in the subspace defined by $z_2=z_{-2}$, where we use the canonical coordinates defined by:
\begin{eqnarray}
q_0&=&Q_0, \;\;p_0=P_0,\nonumber\\
q_2&=&\frac{Q_2+Q_{-2}}{\sqrt{2}},\;\;p_2=\frac{P_2+P_{-2}}{\sqrt{2}}
\end{eqnarray}
These coordinates are related with the real and imaginary part of the complex variable $z_k$, by the following equations:
\begin{eqnarray}
q_0&=&\sqrt{2}u_0,\;\;p_0=\sqrt{2}v_0,\;\;q_2=2u_2,\;\;p_2=2v_2,\\
u_0&=&Re~z_0,\;\;v_0=Im~z_0,\;\;u_2=Re~z_2,\;\;v_2=Im~z_2.\nonumber
\end{eqnarray}
The restriction to the mentioned subspace is justified by the following considerations. It is well known the fact that due to the overcomplete property the coherent states may be used to construct
basis functions describing specific irreducible representations. In particular the coherent state used in this paper can be used to project out the basis $|Nv\alpha JM\rangle$ where the quantum numbers are the boson number $N$, the seniority $v$, the missing quantum number $\alpha$, the angular momentum $J$ and its projection on z-axis $M$ \cite{Rad77}. In Ref.\cite{Rad77} it was shown that a two parameters coherent state is sufficient to be able to project out the entire basis mentioned above. Therefore the properties of the CSM Hamiltonian in the whole boson space can be described in the reduced phase subspace mentioned above. 

In the expression of ${\cal H}$ we adopt the approximation and coordinate transformations:
i) we neglect the terms non-quadratic in momenta as well as the therms coupling the coordinate with momenta; ii)  we take care of the restriction $z_2=z_{-2}$ by introducing the new canonical conjugate coordinates $(q_0;p_0)$ and $(q_2;p_2)$; iii) for the new coordinates we use the polar coordinates:
\begin{equation}
q_0=r\cos\gamma ,\;q_2=r\sin\gamma
\end{equation}   
In this way ${\cal H}$ is a sum of the kinetic and potential energy terms:
\begin{eqnarray}
&&{\cal H}=(11A_1+3A_2+A_1d^2+\frac{3}{70}d^4A_3)(p_0^2+p_2^2)+V(r,\gamma),\nonumber\\
&&V(r,\gamma)=A_1d^4+\frac{A_3}{70}d^6+r^2\left[(11A_1+3A_2)-\frac{d^2}{2}A_1\right.\nonumber\\
&-&\left.\frac{3A_3}{70}d^4-\frac{d^2}{2}A_1\cos(2\gamma)\right]
+\frac{A_3\sqrt{2}}{70}d^3r^3\cos(3\gamma)\nonumber\\
&+&\left(\frac{A_1}{4}+\frac{9A_3}{280}d^2\right)r^4
-\frac{3A_3d}{70\sqrt{2}}r^5\cos(3\gamma)\nonumber\\
&+&\frac{A_3}{280}r^6\left(\cos(6\gamma)+1\right).
\end{eqnarray}

In order to obtain a separable equation for the variables $r$ and $\gamma$ we approximate 
$V(r,\gamma)$ by a sum of two potentials one depending only on $r$, $V_1(r)$, and the other one only on $\gamma$, $V_2(\gamma$). In the terms of $V_1(r)$, the factors depending on $\gamma$ are considered in the minimum point of $V_2(\gamma)$ (which is $\pi/6)$, while in $V_2(\gamma)$ the factors depending on $r$ are considered in the minimum point of $V_1(r)$, denoted by $r_0$. The approximated potential will be denoted by $U(r,\gamma)$. 
\begin{eqnarray}
&&U(r,\gamma)\approx V_1(r)+V_2(\gamma),\nonumber\\
&&V_1(r)=A_1d^4+\frac{A_3}{70}d^6+r^2\left[(11A_1+3A_2)-\frac{3d^2}{4}A_1\right.\nonumber\\
&-&\left.\frac{3A_3}{70}d^4\right]+\left(\frac{A_1}{4}+\frac{9A_3}{280}d^2\right)r^4+\frac{A_3}{280}r^6,\nonumber\\
&&V_2(\gamma)=\frac{A_3}{280}r_0^6\cos(6\gamma).
\end{eqnarray}

In Figs. 1,2, panels (a), (c), (e) we present the contour plot and two sections of the approximate potential $U(r,\gamma)$. The two sections are obtained by fixing one of the two variables in the potential minimum point.  In the panels (b), (d) and (e) similar plots for the exact potential $V(r,\gamma)$ are given. Figures 1 and 2 correspond to different sets of the CSM parameters, i.e.
$d, A_1, A_2$ and $A_3$. Comparing the curves from left and right panels of each figure we conclude that the effect of approximations yielding the separated form of the potentials
is quite small. The only visible effect is on the orientation of the symmetry axis of the equipotential curves.
Indeed, the contour plots for exact and approximated potentials  have different symmetry axes although they surround the same minimum point. Changing slightly the CSM parameters we note that indeed the curves from Fig. 2 are quite close to the corresponding curves from Fig.1. Thus, we may say that indeed, the curves are stable against small changes of the set of parameters determining the classical $r$ and $\gamma$ potential. 
\begin{figure}[h!]
\begin{center}
\includegraphics[width=8.5cm]{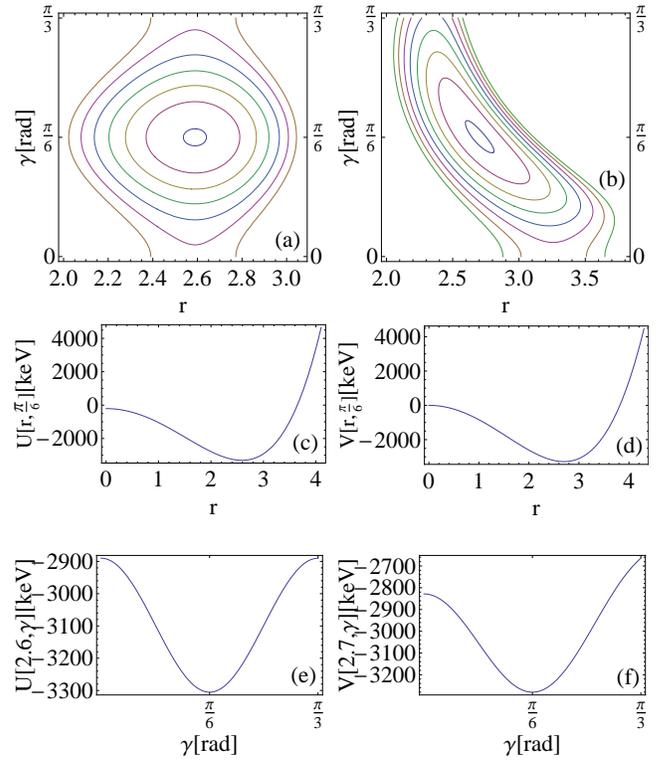}
\caption{(Color online) The contour plot for the approximated potential $U(r,\gamma)$, given in (4.14), is presented in the variables $r$ and $\gamma$ (a). The parameters involved are $ d=3.5,  A_{1}=10.65 keV$, $A_{2}=15.146 keV$, and $A_{3}=150 keV$. The minimum is reached in 
$(r_{0},\gamma_0)=(2.6,\pi/6)$.
The equipotential curves are separated by an amount of 80 keV.
The minimum value of the potential is $U_{min}=-3305.25 keV$. Two sections $\gamma=\frac{\pi}{6}$ and $r_0=2.6$ of the potential $U(r,\gamma)$ are presented in panels (c) and (e), respectively.
Similar plots but for the exact potential from (4.13) are given in the panels (b), (d) and (f),
respectively. Parameters were kept the same as for the approximated potential.  The minimum value of the potential is $V_{min}=-3280.31 keV$ and is reached at $(r_{0},\gamma_0)=(2.7,\pi/6)$..The equipotential curves are separated by an amount of 50 keV.}
\end{center}
\label{Fig. 1}
\end{figure}

\begin{figure}[h!]
\begin{center}
\includegraphics[width=8.5cm]{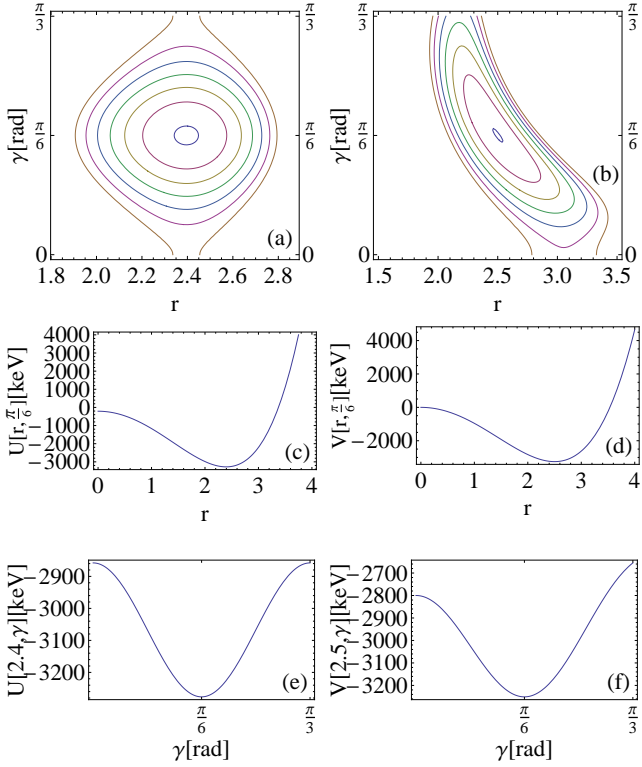}
\caption{(Color online) The contour plot for the approximated potential $U(r,\gamma)$ given by Eq. (4.14) is represented as function of the variables $r$ and $\gamma$ (a). The parameters involved are $ d=3.22,  A_{1}=12.95 keV$, $A_{2}=7.9keV$, and $A_{3}=240 keV.$
 The minimum is reached in 
$(r_{0},\gamma_0)=(2.4,\frac{\pi}{6})$.
The equipotential curves surrounding the minimum value are separated by an amount of 70 keV.
The minimum value of the potential is $U_{min}=-3276.43 keV$. Two sections $\gamma=\frac{\pi}{6}$ and $r_0=2.4$ of the potential $U(r,\gamma)$ are presented in the panels (c) and (e), respectively.
 Similar plots but for the exact potential from (4.13) are given in the panels (b), (d) and (f),
respectively. Parameters were kept the same as for the approximated potential. The minimum value of the potential is $V_{min}=-3250.92 keV$ and is reached at $(r_{0},\gamma_0)=(2.5,\frac{\pi}{6})$. The equipotential curves are separated by an amount of 50 keV.}
\end{center}
\label{Fig. 2}
\end{figure}

The classical Hamilton function becomes:
\begin{equation}
{\cal H}=(11A_1+3A_2+A_1d^2+\frac{3}{70}d^4A_3)(p_0^2+p_2^2)+U(r,\gamma).
\end{equation}
This can be quantized by replacing the sum of the momenta squared by the Laplace operator written in polar coordinates:
\begin{eqnarray}
\hat{H}&=&-(11A_1+3A_2+A_1d^2+\frac{3}{70}d^4A_3)\\
        &\times&\left(\frac{1}{r}\frac{\partial}{\partial r}+
\frac{\partial^2}{\partial r^2}+\frac{1}{r^2}\frac{\partial^2}{\partial \gamma^2}\right)
+V_1(r)+V_2(\gamma).\nonumber
\end{eqnarray}
It is convenient to introduce the notation:
\begin{equation}
{\cal F}=(11A_1+3A_2+A_1d^2+\frac{3}{70}d^4A_3).
\end{equation}
The  Schr\"{o}dinger equation:
\begin{equation}
\hat{H}\Psi(r,\gamma)=E\Psi(r,\gamma),    
\end{equation}
for the trial function:
\begin{equation}
\Psi(r,\gamma)=\psi_1(r)e^{iK\gamma}\psi_2(\gamma),
\end{equation}
is separated:
\begin{eqnarray}
&&\left[-{\cal F}\left(\frac{1}{r}\frac{\partial}{\partial r}+
\frac{\partial^2}{\partial r^2}-\frac{K^2}{r^2}\right)+V_1(r)\right]\psi_1(r)=
E^{(1)}\psi_1(r),\nonumber\\
&&\left[-{\cal F}\left(\frac{2iK}{r_0^2}\frac{\partial }{\partial\gamma}+\frac{1}{r_0^2}\frac{\partial^2}{\partial \gamma^2}\right)+V_2(\gamma)\right]\psi_2(\gamma)=E^{(2)}\psi_2(\gamma).\nonumber\\
\label{SMA}
\end{eqnarray}
In what follows we shall show that the first equation (\ref{SMA}) leads to a Schr\"{o}dinger equation for a sextic potential plus a centrifugal term, while the second equation provides a differential equation obeyed by the Mathieu function.

Dividing both sides of the first Eq.(\ref{SMA}) by ${\cal F}$ and denoting by
\begin{equation}
u_{1}(r)=V_{1}(r)/{\cal F},\;\; \varepsilon_{r}=E^{(1)}/{\cal F}
\end{equation}
the equation in the variable $r$ becomes:
\begin{equation}
\left[-\frac{\partial^{2}}{\partial r^{2}}-\frac{1}{r}\frac{\partial}{\partial r}+\frac{K^{2}}{r^{2}}+u_{1}(r)\right]\psi_{1}(r)=\varepsilon_{r}\psi_{1}(r).
\end{equation}
Changing the function 
\begin{equation}
\psi_{1}(r)=r^{-\frac{1}{2}}\phi(r),
\end{equation}
the  equation for the new function is 
\begin{equation}
\left[-\frac{\partial^{2}}{\partial r^{2}}+\frac{K^{2}-\frac{1}{4}}{r^{2}}+u_{1}(r)\right]\phi(r)=\varepsilon_{r}\phi(r),
\end{equation}
which is nothing else but the Schr\"{o}dinger equation for a sextic potential plus a centrifugal term. 

Now let us turn our attention to the second equation from (\ref{SMA}).
Multiplying it with $r_0^2/{\cal F}$ and denoting by:
\begin{equation}
\mu=\frac{A_{3}}{280{\cal F}}r_{0}^{8}, \;\;\varepsilon_{\gamma}=\frac{r_{0}^{2}E^{(2)}}{{\cal F}},
\end{equation}
one obtains:
\begin{equation}
\left[-\frac{\partial^{2}}{\partial\gamma^{2}}-2iK\frac{\partial}{\partial\gamma}+\mu\cos6\gamma\right]\psi_{2}(\gamma)=\varepsilon_{\gamma}\psi_{2}(\gamma).
\end{equation}
With the change of function:
\begin{equation}
\psi_{2}(\gamma)=e^{-iK\gamma}M(3\gamma),
\end{equation}
we obtain:
\begin{equation}
\left[\frac{\partial^{2}}{\partial\gamma^{2}}+\varepsilon_{\gamma}+K^{2}-\mu\cos6\gamma\right]M(3\gamma)=0.
\end{equation}
Changing now the variable $\gamma$ to $y=3\gamma$, the equation for the Mathieu function is readily obtained:
\begin{equation}
\left[\frac{\partial^{2}}{\partial y^{2}}+a-2q\cos2y\right]M(y)=0,
\end{equation}
where the following notations have been used:
\begin{equation}
a=\frac{1}{9}(\varepsilon_{\gamma}+K^{2}),\hspace{0.5cm}2q=\frac{\mu}{9}.
\end{equation}

Before closing this section we would like to comment on the relation of the variable $r$ and the dynamic nuclear deformation $\beta$. 

Aiming at this goal let us consider the canonical transformation relating the quadrupole conjugate 
coordinates and the boson operators:

\begin{eqnarray}
\hat{\alpha}_{2\mu}&=&\frac{1}{k\sqrt{2}}\left(b^{\dagger}_{2\mu}+(-)^{\mu}b_{2,-\mu}\right),\nonumber\\
\hat{\pi}_{2\mu}&=&\frac{ik}{\sqrt{2}}\left(b^{\dagger}_{2,-\mu}(-)^{\mu}+b_{2\mu}\right).
\end{eqnarray}
Note that the canonical transformation from above is determined up to a multiplicative factor $k$.
Averaging these equations with the coherent state $\psi$ (5.1), one obtains that the coordinates 
$Q_{\mu}$ and $P_{\mu}$ introduced above are related with the quadrupole operators by:
\begin{equation}
Q_{\mu}=\langle \psi|k\hat{\alpha}_{2\mu}|\psi \rangle,\;\;
P_{\mu}=\langle \psi|\frac{1}{k}\hat{\pi}_{2\mu}|\psi \rangle.
\end{equation}
Identifying the averages of $\hat{\alpha}_{2\mu}$ with the coherent state, with the quadrupole coordinates in the intrinsic reference frame we obtain \cite{RaBaDe}:
\begin{equation}
Q_{0}=ka_{20}=k\beta\cos\gamma,\;\;Q_{\pm2}=\frac{k\beta}{\sqrt{2}}\sin\gamma .
\end{equation}
In the restricted classical phase space the canonical coordinates are:
\begin{equation}
q_0=k\beta\cos\gamma,\;\; q_2=k\beta\sin\gamma.
\end{equation}
From here it results:
\begin{equation}
r=k\beta
\end{equation}
Using this simple relation in connection with the differential equation in $r$, one obtains the Shr\"{o}dinger equation for sextic potential plus centrifugal term in the variable $\beta$.
\begin{equation}
\left[-\frac{\partial^{2}}{\partial \beta^{2}}+\frac{K^{2}-\frac{1}{4}}{\beta^{2}}+k^2u_{1}(k\beta)\right]\phi(k\beta)=\varepsilon_{\beta}\phi(k\beta)
\end{equation}
Concluding this section, we may say that while in the previous paper \cite{RaBu011}
the sextic potential for $\beta$ and the $\gamma$ potential yielding the equation for the Mathieu function were introduced by an ad-hoc choice, here they are derived in a natural manner from the CSM formalism.  Moreover the variable separation is based on two approximations suggested by the classical picture: i) the non-quadratic terms in momenta are ignored and ii) the coupling of coordinates and momenta are vanishing due to the local character of the classical phenomenological forces.

\renewcommand{\theequation}{5.\arabic{equation}}
\setcounter{equation}{0}
\section{Numerical results}
\label{sec:level5}
The formalisms SMA and CSM presented in the previous sections have been applied for calculating the excitation energies and the available B(E2) values for two isotopes:$^{180}$Hf, $^{182}$W. We start with the excitation energy analysis.
As shown above the total energy provided by SMA depends on five parameters:
$\hbar^2/2B, a, b, \frac{1}{\langle \beta^2\rangle},\mu$. These have been fixed by fitting the
excitation energies using the least square procedure. The results are given in Table I.
Concerning CSM, the parameters determining the energies are: $d,A_1,A_2,A_3$. They were fixed as follows. We cycled $d$ within a large interval with a small step. For each $d$ we determined $A_1$ and $A_2$ by fitting the energies of two states, one belonging to the ground band and one from the gamma band. $A_3$ was obtained by fitting one level energy from the beta band. Then we choose that $d$ which yields an overall good fit. The fitted parameters are given in Table I.
\begin{table}[b!]
\begin{tabular}{|c|c|c|}
  \hline
   & $^{180}Hf$ & $^{182}W$  \\
   \hline
  $\frac{\hbar^{2}}{2B}[keV]$& 0.401& 0.476888  \\
  \hline
  $a$ & 16212.86 & 11508.56 \\
  \hline
  $b$ & -44 & 95 \\
  \hline
  $\frac{1}{\langle\beta^{2}\rangle}$ & 3.14 & 3.457\\
  \hline
  $\mu$ & 20892. & 12772.6 \\
  \hline
  $t_{1}$& 177.93[W.u]$^{1/2}$ & 158.48[W.u]$^{1/2}$  \\
  \hline
  $t_{2}$& 4630.24[W.u.]$^{1/2}$ & 4736.037 [W.u.]$^{1/2}$ \\
  \hline
  d         &3.5     &3.22     \\
  \hline 
  $A_1$[keV]&21.17   & 21.54 \\
  \hline
  $A_2$[keV]&8.15   & 7.47 \\
  \hline
  $A_3$[keV]&-10.85  & -12.29  \\
  \hline
  $q_1$& 3.739[W.u.]$^{1/2}$&3.756[W.u.]$^{1/2}$\\           
  \hline
   $q_2$&-0.125[W.u.]$^{1/2}$&-0.175[W.u.]$^{1/2}$\\
  \hline  
\end{tabular}
\caption{The parameters $\hbar^{2}/2B$, $a$, $b$, $\frac{1}{\langle\beta^{2}\rangle}$, $\mu$ involved in the energy expression provided by SMA (\ref{totalen}), are given for $^{180}$Hf and 
$^{182}$W. Also we give the values for the parameters $t_{1}$ and $t_{2}$ defining the transition operator used by SMA (\ref{tranoper}). On the last six rows we give the parameters determining the CSM excitation energies, $d, A_1, A_2, A_3$, and the specific E2 transition operator i.e., $q_1$ and $q_2$. }
\end{table}

The results obtained with the two approaches, $SMA$ and $CSM$, are compared, in Fig.3 and Fig.4,
with the corresponding experimental data.

\begin{figure}[h!]
\begin{center}
\includegraphics{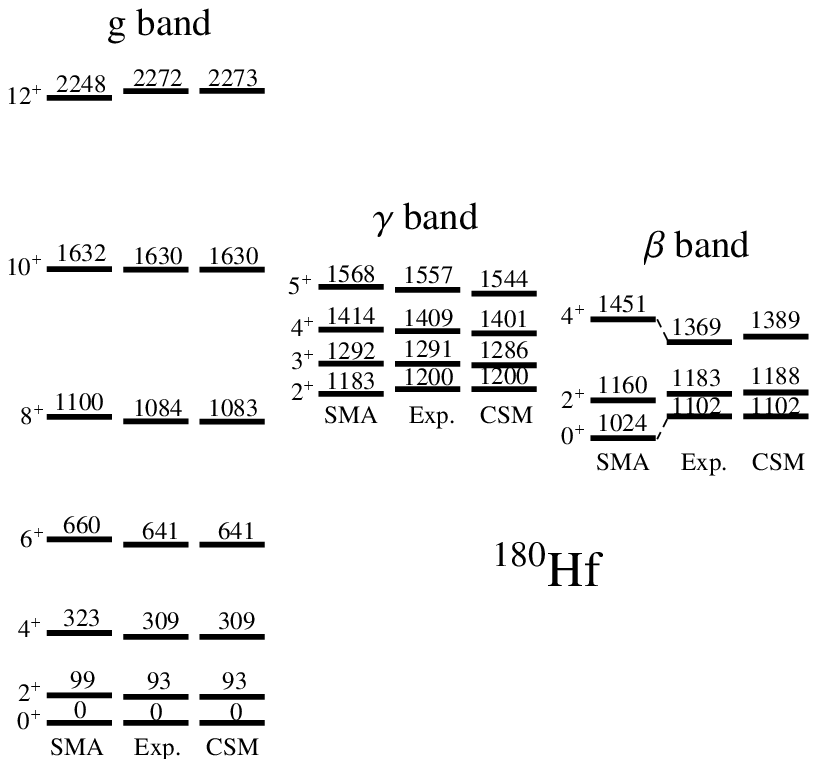}
\end{center}
\caption{Excitation energies for ground, beta and gamma bands in $^{180}$Hf, obtained with SMA and CSM formalism
respectively, are compared with the corresponding experimental data taken from Ref.\cite{Wu}.}
\label{Fig. 5}
\end{figure}
\begin{figure}[h!]
\begin{center}
\includegraphics{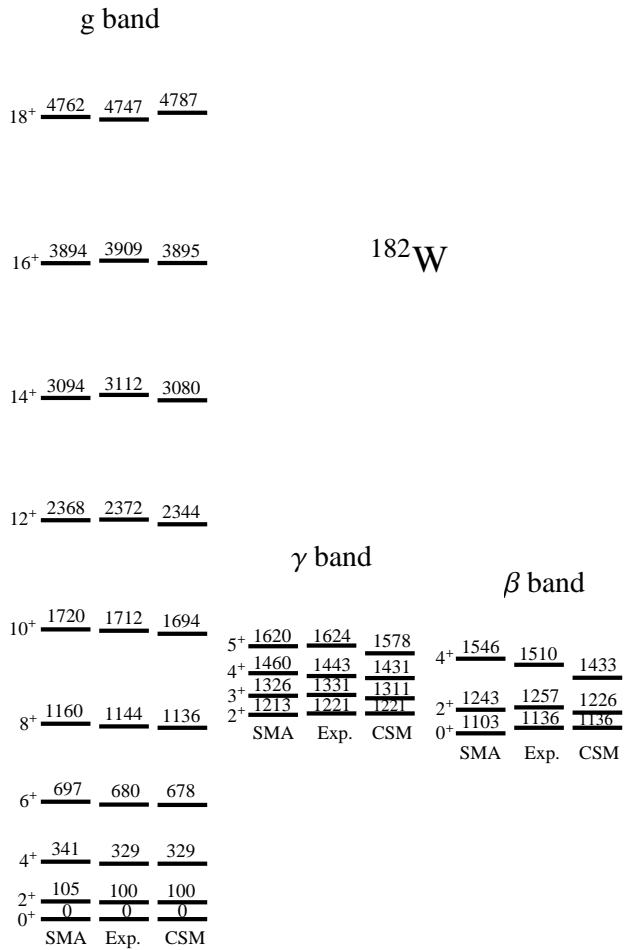}
\end{center}
\caption{Excitation energies for ground, beta and gamma bands in $^{182}$W, obtained with SMA and CSM formalism
respectively, are compared with the corresponding experimental data taken from Ref.\cite{Bal}.}
\label{Fig. 6}
\end{figure}

By inspecting the experimental data we may see whether the signature for triaxiality shows up.
Indeed, it is well known the fact that the most distinctive signature of the triaxial rigid rotor is the equation relating the energies of three particular states \cite{Filip,Chaban,Davydov}: 
\begin{equation}
E_{2^+_1}+E_{2^+_2}=E_{3^+_1}.
\label{Davydeq}
\end{equation}
Actually this equation is only approximately obeyed. Denoting by $\Delta E$ the modulus of the difference between the left and right hand side of the mentioned relation, the experimental data  lead to the values:
\begin{equation}
\Delta E=2 \rm{keV};10 \rm{keV}.
\end{equation}
for $^{180}$Hf and $^{182}$W, respectively.
Clearly, these deviations suggest that the nuclei considered in the present paper are close to 
an ideal triaxial rotor. As a matter of fact this is the experimental feature which inspired us to take the $\gamma=30^0$ as  reference picture. 

Let us  discuss now the results concerning the transition probabilities.
The SMA made use of a anharmonic transition operator written in the intrinsic 
frame of reference (\ref{tranoper}), while CSM employs a second order boson operator in the laboratory frame (\ref{q2anh}). In both cases the operators involve two parameters: $t_1$ and $t_2$
for SMA and $q_1$ and $q_2$ for CSM. These parameters have been fixed by fitting two particular transitions for each nucleus. The fitted parameters are given in Table I.
Results of our calculations and the corresponding experimental data are given in Table II for $^{180}$Hf and  Table III for $^{182}$W. 

We remark that both formalisms describe reasonable well the reduced transition probabilities.
It is worth noting the high accuracy of the CSM description of both excitation energies and transition probabilities for $^{180}$Hf. 
\begin{table}[h!]
\begin{tabular}{|c|c|c|c|}
\hline
                       &\multicolumn{3}{|c|}{$^{180}$Hf}\\
\cline{2-4}
B(E2;$J_i^+\to J_f^+$)                       & SMA    &   Exp.   &   CSM   \\
\hline
$2_g^+\to 0_g^+$         & 155    &    155   & 155      \\  
$4_g^+\to 2_g^+$         & 219    &    230   & 223       \\   
$6_g^+\to 4_g^+$         & 281    &    219   & 250       \\
$8_g^+\to 6_g^+$         & 321    &    250   & 267       \\ 
$10_g^+\to 8_g^+$        & 353    &    240   & 282        \\
$12_g^+\to 10_g^+$       & 380    &    232   & 296       \\  
$2_{\gamma}^+\to 0_g^+$  & 0.2    &    3.8   & 3.8       \\ 
$2_{\gamma}^+\to 2_g^+$  & 5.1    &    5.1   & 6.6        \\
\hline
\end{tabular}
\caption{ B(E2) values for some ground to ground and gamma to ground E2 transitions
in $^{180}$Hf. $J_i^+$ and $J_f^+$ denote the angular momenta of the initial and final states, respectively. Experimental data were taken from Ref.\cite{Wu}}
\end{table}

Before closing this section we would like to spend  few words about the other descriptions of the triaxiality features. We have not to forget however the fact that the pioneering paper for 
phase transition from gamma stable to gamma unstable nuclei with an analytical description of the critical point, and the departure from axial symmetry
is that of Jean-Wilets \cite{Wilet}. However an intensive study of the subject started in the beginning of the last decade, since the context of symmetries was much developed and on the other hand relevant data have been accumulated.
Triaxiality has been investigated within the IBA formalism being related to various effects.
Thus including higher order terms the triaxiality of $^{190,192}$Os has been studied in Ref.
\cite{Tiamova}. Including the $g$-boson, in a recent study \cite{VanIsacker} no shape/phase transition towards stable triaxial shapes has been found. The phase diagram of IBA-2 (which distinguishes protons from neutrons) including triaxial shapes, has been constructed in Refs.
\cite{Arias,Caprio1,Caprio2}. 
\begin{table}[t!]
\begin{tabular}{|c|c|c|c|}
\hline
                       &\multicolumn{3}{|c|}{$^{182}$W}\\
\cline{2-4}
B(E2;$J_i^+\to J_f^+$)                        & SMA    &   Exp.   &   CSM   \\
\hline
$2_g^+\to 0_g^+ $        & 137    &137       & 137      \\  
$4_g^+\to 2_g^+ $        & 194    &196      &  198      \\   
$6_g^+\to 4_g^+ $        & 248    &200       & 223       \\
$8_g^+\to 6_g^+ $        & 284    &209      &  241     \\ 
$10_g^+\to 8_g^+$        & 312    &203      &  256       \\
$12_g^+\to 10_g^+$       & 336    &191       & 271       \\   
$14_g^+\to 12_g^+$       & 358    &170       & 285       \\   
$16_g^+\to 14_g^+$       & 379    &204       & 300       \\   
$18_g^+\to 16_g^+$       & 398    &250       & 315       \\   
$2_{\beta}^+\to 0_{\beta}^+$&117  &200       & 157      \\
$2_{\beta}^+\to 0_{g}^+$&1.3      &0.9       & 0.008      \\
$2_{\beta}^+\to 4_{g}^+$&10.5     &1.7       & 0.021      \\
$2_{\gamma}^+\to 0_g^+ $&0.2      &3.4       & 3.4      \\
$2_{\gamma}^+\to 2_g^+ $&8.5      &6.74      & 6.27      \\
$2_{\gamma}^+\to 4_g^+ $&0.0      &0.034     & 0.51      \\
$4_{\gamma}^+\to 2_g^+ $&0.1      &2.4       & 1.36       \\ 
$4_{\gamma}^+\to 4_g^+ $&1.7      &10.4      & 7.60       \\
\hline
\end{tabular}
\caption{B(E2) values for some ground to ground, beta to ground and gamma to  ground E2 transitions
in $^{182}$W. $J_i^+$ and $J_f^+$ denote the angular momenta of the initial and final states, respectively. Experimental data were taken from Ref.\cite{Bal}.}
\end{table}

Several authors treated the gamma soft nuclei around $\gamma^0=\frac{\pi}{6}$ \cite{Bona1,fortu1,fortu2,fortu3}. However, their equations for beta as well for gamma variables are different from those proposed in the present paper. The beta potential is either an infinite square well \cite{Bona1} or a Coulomb or a Kratzer potential \cite{fortu1,fortu2,fortu3}. Recently \cite{Yigitoglu}, the Davidson potential was used in relation to triaxial nuclei. Concerning $\gamma$ all quoted descriptions use an oscillator potential. The sextic potential was previously used in Refs.\cite{Levai,Levai1} but only for few
low lying states. Again the description of $\gamma$ is different.
Triaxiality has recently been studied in the framework of the algebraic collective model 
\cite{Rowe}, and the onset of rigid triaxial deformation has been considered \cite{Caprio3}.

The structure of the projected states in the intrinsic variable is relevant for the subject under consideration. In Ref. \cite{RadKopel} we studied the probability distributions of the $\gamma$ and the $\beta$ variables corresponding to various states from the ground, beta and gamma bands. For the state $0^+_{\beta}$ the $\gamma$ probability has a minimum in $\pi/6$ and two maxima in $\gamma=0$ and 
$\gamma=\pi/3$, respectively. Increasing the spin in the $\beta$ band, the picture is changed. For example for 
$10^+_{\beta}$ the probability distribution of $\gamma$ has a minimum  in $\pi/6$ and two maxima, one for $\gamma =\pi/12$ and one for $\gamma=\pi/4$. The $\gamma$ band exhibits an oposite feature, namely the head state has a maximum in $\pi/6$, while in a high state, like $J^{\pi}=10^+$, two proeminent maxima for $\gamma=0$ and $\gamma=\pi/3$ are observed. In the ground state $0^+_g$ one meet a situation which  is specific to gamma unstable nuclei, i.e. the distribution is almost constant.
Increasing the spin one obtains for $J^{\pi}=10^+$ a maximum at $35^0$. In beta variable the state $0^+_{\beta}$ has a bimodal structure, the head state of gamma band has a deformed maximum while the ground state a spherical maximum. If the factor $\beta^4$, entering the  measure of integration over the dynamical deformation $\beta$, then the distributions are peaked on various deformations, depending on the band to which the state belongs and of course on the angular momentum.  The resuts mentioned above are consistent with the predictions of the semiphenomenological model proposed by Kumar and Baranger \cite{Baranger}. 

Recall the fact that the ground band states with $J\ge 2$ are mixed with the states of the same angular momentum from gamma band. Therefore one expects to have  gamma distributions with a maximum value at $\gamma=\pi/6$,
at least for low lying states from the ground and gamma bands, while in the beta band two maxima at about $\pi/12$ and $\pi/4$ may show up.

The transition from a near spherical shape to a triaxial shape is accompanied by the change of the staggering from $2^+, (3^+, 4^+), (5^+,6^+),...$ to $(2^+,3^+), (4^+,5^+), (6^+,7^+),...$,
In Ref. \cite{RaSa} we showed that whithin CSM the first clustering is remnant of the doublet degeneracies of the vibrational limit of the model, while the second staggering which is typical for rigid triaxial rotor, is obtained in the asymmptotic region of deformation. Also, such a transition is reflected in the behaviour of the E2 transitions in the gamma band.

We remember that SMA was initially \cite{RaBu011} obtained from the Bohr-Mottelson Hamiltonian
by separating the variables $\beta$ and $\gamma$ and supplementing the result with new potentials for beta and gamma respectively. Here SMA is derived from CSM based on a similar procedure of the variable separation. Of course it would be nice if the final quasiexactly separable approach keeps track of the formalism from which it emerges. This problem was studied in great details by Caprio \cite {Caprio05} in connection with the X(5) Hamiltonian comparing the results of approximate separated Hamiltonian with the exact ones obtained by diagonalizing the initial Hamiltonian in a large  basis constructed from a five dimensional spherical harmonics
\cite{Ghe,Rad78,Row1,Row2,Row3}. The conclusion was that replacing $1/\beta^2$ by the "rigid" value
$1/\beta_0^2$ is a valid approximation for nearly $\gamma$ soft potential, while the "small" angle approximation for $\gamma$ is good for large  $\gamma$ stiffnesses. Both approximations work in the overlapping region of the two $\gamma$ intervals.

Within the SMA the situation is completely different. The only term approximated in the $\gamma$
equation is $1/\sin^23\gamma$ originating from the $\gamma$ kinetic energy term. However, this is not expanded around $\gamma=0$ which is related to the $\gamma$- small picture but around $\pi/6$ where the mentioned term is minimum. In Ref.\cite{RaBu011} we studied the potentials in $\gamma$ for several nuclei and noticed that they are characterized by a large stiffness.
Concerning the approximation in $\beta$, in the quoted paper we plotted the functions of different states from the three bands. For all considered  cases, the functions are highly localized on $\beta_0$. There is an esential difference between our case and that of Ref.\cite{Caprio05}. Indeed, there the Hamiltonian which is separated is harmonic and the approximations concern the 
part of centrifugal term $1/\beta_0^2$ which is distributed to the factor expressing the $\gamma$
kinetic energy and the quadratic Taylor expansion around $\gamma=0$ giving the fact that this is a singularity point for the $\gamma$ potential. Note that in our descrition the Hamiltonian is highly anharmonic and moreover the $\gamma$ expansion is made around a stability point for the
$\gamma$ potential. Moreover the $\beta$ potential has a deep deformed minimum which favor a small contribution of fluctuations which might shift the average $\langle\beta^2\rangle$ from $\beta_0^2$.

Concerning the SMA derivation from the CSM, we don't have a centrifugal term in $r$ which is coupled to the terms from the $\gamma$ kinetic energy, and therefore the trouble generated by replacing
$r$ with $r_0$ in a centrifugal like term is absent. However, the separation is ultimately obtained by replacing $\gamma$ by the minimum point coordinate $\gamma_0$ in some terms, and $r$ with 
the minimum coordinate $r_0$ in other terms. Such an aproximation is justified for situations when the eigenfunctions of the $\gamma$ and $r$ Hamiltonian respectively, are well centered on the 
$\gamma_0$ and $r_0$. This picture might be reached indeed, since as shown in Figs. 1, 2, the potentials depths are large. Actually, a confident measure of the separation approximation quality is the comparison of the
approximated and non-approximated potentials shown in Figs 1, 2 for two sets of parameters.
As seen there, the potentials do not differ from each other too much as to induce a significant difference of the corresponding wave function.

\clearpage

\renewcommand{\theequation}{6.\arabic{equation}}
\setcounter{equation}{0}
\section{Conclusions}
Here we summarize the main results obtained in the previous sections. The formalism,  conventionally called {\bf S}extic and {\bf M}athieu {\bf A}pproach (SMA), proposed in a previous publication for the description of the triaxial nuclei is applied to describe the spectra and transition probabilities for another two nuclei, $^{180}$Hf and $^{182}$W. Results of our calculations are in good agreement with the available data.

{\it The main result of the present study is a natural derivation of the SMA as a limiting case of the CSM.} The model Hamiltonian of CSM together with a three parameters quadrupole coherent state is used in a time dependent variational principle to derive the classical equations of motion in the classical phase space coordinates. In the classical energy function one neglects the terms which are non-quadratic in momenta as well as those which couple coordinates and momenta. Reducing the space to the subspace generated by two pairs of conjugate coordinates and then quantizing the classical Hamilton function one arrives at a separable form for the associated Schr\"{o}dinger equation, one being the equation for a sextic oscillator in $\beta$ and the other one a differential equation for $\gamma$ obeyed by the Mathieu function. The reduced space corresponds to a two parameters coherent state from which the entire boson space basis
$|Nv\alpha JM\rangle$ can be projected out. This justifies the restriction since in this way one accounts for the classical properties of the CSM model Hamiltonian in the whole boson space.

The final conclusions are:
\begin{itemize}
\item{The sextic potential for the $\beta$ variable  and the potential in $\gamma$, introduced in our previous paper on this subject 
\cite{RaBu011}, based on pragmatic grounds, gets a theoretical support.}
\item{The success of CSM in explaining the data in a realistic way lets us conclude that, indeed,  this model is able to describe the triaxial nuclei.}
\item{ According to the results of the present paper, the SMA represents the image of CSM in the intrinsic reference frame.} 
\item{We stress, again, that in describing the near vibrational, well deformed with axial symmetry, triaxial or gamma unstable nuclei one uses a sole model Hamiltonian and a sole set of model functions for the states belonging to the ground, beta and gamma bands.}
\item{ This makes us hope that the phase transition from prolate to oblate shape through the triaxial shape can be described by CSM.}
\end{itemize}
This feature will be however presented, soon,  in another publication.

{\bf Acknowledgment.} This work was supported by the Romanian Ministry for Education Research Youth and Sport through the CNCSIS project ID-1038/2008.

\renewcommand{\theequation}{A.\arabic{equation}}
\setcounter{equation}{0}
\section{Appendix A}

\begin{eqnarray}
&&H=(11A_{1}+3A_{2})(P_{0}^{2}+2P_{-2}P_{2}+Q_{0}^{2}+2Q_{-2}Q_{2})\nonumber\\
 &+&A_{1}(d^{4}+d^{2}P_{0}^{2}+\frac{P_{0}^{4}}{4}+2d^{2}P_{-2}P_{2}+P_{-2}P_{0}^{2}P_{2}\nonumber\\
&+&P_{-2}^{2}P_{2}^{2}+P_{-2}^{2}Q_{-2}^{2}+2P_{-2}P_{0}Q_{-2}Q_{0}-d^{2}Q_{0}^{2}\nonumber\\
&+&\frac{1}{2}P_{0}^{2}Q_{0}^{2}-P_{-2}P_{2}Q_{0}^{2}+\frac{Q_{0}^{4}}{4}-2d^{2}Q_{-2}Q_{2}-P_{0}^{2}Q_{-2}Q_{2}\nonumber\\
&+&2P_{0}P_{2}Q_{0}Q_{2}+Q_{-2}Q_{0}^{2}Q_{2}+P_{2}^{2}Q_{2}^{2}+Q_{-2}^{2}Q_{2}^{2})\nonumber\\
&+&\frac{A_{3}}{70}(d^{6}+3d^{4}P_{0}^{2}+\frac{9}{4}d^{2}P_{0}^{4}+\frac{P_{0}^{6}}{2}+6d^{4}P_{-2}P_{2}\nonumber\\
&+&9d^{2}P_{-2}P_{0}^{2}P_{2}-6P_{-2}P_{0}^{4}P_{2}+9d^{2}P_{-2}^{2}P_{2}^{2}\nonumber\\
&+&18P_{-2}^{2}P_{0}^{2}P_{2}^{2}+6\sqrt{2}d^{3}P_{-2}P_{0}Q_{-2}+12\sqrt{2}d P_{-2}P_{0}^{3}Q_{-2}\nonumber\\
&+&9d^{2}P_{-2}^{2}Q_{-2}^{2}+18P_{-2}^{2}P_{0}^{2}Q_{-2}^{2}-3\sqrt{2}d^{3}P_{0}^{2}Q_{0}\nonumber\\
&-&\frac{3d P_{0}^{4}Q_{0}}{\sqrt{2}}+6\sqrt{2}d^{3}P_{-2}P_{2}Q_{0}-18\sqrt{2}d P_{-2}P_{0}^{2}P_{2}Q_{0}\nonumber\\
&+&18\sqrt{2}dP_{-2}^{2}P_{2}^{2}Q_{0}+18d^{2}P_{-2}P_{0}Q_{-2}Q_{0}-12P_{-2}P_{0}^{3}Q_{-2}Q_{0}\nonumber\\
&+&18\sqrt{2}dP_{-2}^{2}Q_{-2}^{2}Q_{0}-3d^{4}Q_{0}^{2}+\frac{9}{2}d^{2}P_{0}^{2}Q_{0}^{2}\nonumber\\
&+&\frac{3}{2}P_{0}^{4}Q_{0}^{2}-9d^{2}P_{-2}P_{2}Q_{0}^{2}+18P_{-2}^{2}P_{2}^{2}Q_{0}^{2}\nonumber\\
&+&18P_{-2}^{2}Q_{-2}^{2}Q_{0}^{2}+\sqrt{2}d^{3}Q_{0}^{3}-3\sqrt{2}d P_{0}^{2}Q_{0}^{3}-6\sqrt{2}dP_{-2}P_{2}Q_{0}^{3}\nonumber\\
&-&12P_{-2}P_{0}Q_{-2}Q_{0}^{3}+\frac{9}{4}d^{2}Q_{0}^{4}+\frac{3}{2}P_{0}^{2}Q_{0}^{4}+6P_{-2}P_{2}Q_{0}^{4}\nonumber\\
&-&\frac{3dQ_{0}^{5}}{\sqrt{2}}+\frac{Q_{0}^{6}}{2}+6\sqrt{2}d^{3}P_{0}P_{2}Q_{2}+12\sqrt{2}d P_{0}^{3}P_{2}Q_{2}\nonumber\\
&-&6d^{4}Q_{-2}Q_{2}-9d^{2}P_{0}^{2}Q_{-2}Q_{2}+6P_{0}^{4}Q_{-2}Q_{2}\nonumber\\
&+&18d^{2}P_{0}P_{2}Q_{0}Q_{2}-12P_{0}^{3}P_{2}Q_{0}Q_{2}-6\sqrt{2}d^{3}Q_{-2}Q_{0}Q_{2}\nonumber\\
&+&18\sqrt{2}dP_{0}^{2}Q_{-2}Q_{0}Q_{2}+9d^{2}Q_{-2}Q_{0}^{2}Q_{2}-12P_{0}P_{2}Q_{0}^{3}Q_{2}\nonumber\\
&+&6\sqrt{2}dQ_{-2}Q_{0}^{3}Q_{2}-6Q_{-2}Q_{0}^{4}Q_{2}+9d^{2}P_{2}^{2}Q_{2}^{2}\nonumber\\
&+&18P_{0}^{2}P_{2}^{2}Q_{2}^{2}+9d^{2}Q_{-2}^{2}Q_{2}^{2}+18P_{0}^{2}Q_{-2}^{2}Q_{2}^{2}+18\sqrt{2}dP_{2}^{2}Q_{0}Q_{2}^{2}\nonumber\\
&+&18\sqrt{2}dQ_{-2}^{2}Q_{0}Q_{2}^{2}+18P_{2}^{2}Q_{0}^{2}Q_{2}^{2}+18Q_{-2}^{2}Q_{0}^{2}Q_{2}^{2}).
\end{eqnarray}

\end{document}